\documentclass[prl,aps,twocolumn,superscriptaddress,showpacs,preprintnumbers,amssymb]{revtex4}

\usepackage{color}
\usepackage{graphicx}
\usepackage{ulem}
\usepackage{float}
\usepackage{here}
\usepackage{bm}
\usepackage{amstext}
\usepackage{amsfonts}
\usepackage{dcolumn} 
\usepackage[colorlinks=true,urlcolor={blue},citecolor={blue}]{hyperref}
\usepackage{pgf}
\usepackage{graphicx}
\usepackage[T1]{fontenc}

\def\textbf#1{\boldsymbol{#1}}

\begin{document}

\title{Rapid Suppression of the Charge Density Wave in YBa$_2$Cu$_3$O$_{6.6}$ under Hydrostatic Pressure}
\author{S. M. Souliou}
\affiliation{European Synchrotron Radiation Facility, BP 220, F-38043 Grenoble Cedex, France}
\affiliation{Max-Planck-Institut f\"{u}r Festk\"{o}rperforschung, Heisenbergstrasse 1, D-70569 Stuttgart, Germany}
\author{H. Gretarsson}
\affiliation{Max-Planck-Institut f\"{u}r Festk\"{o}rperforschung, Heisenbergstrasse 1, D-70569 Stuttgart, Germany}
\author{G. Garbarino}
\affiliation{European Synchrotron Radiation Facility, BP 220, F-38043 Grenoble Cedex, France}
\author{A. Bosak}
\affiliation{European Synchrotron Radiation Facility, BP 220, F-38043 Grenoble Cedex, France}
\author{J. Porras}
\affiliation{Max-Planck-Institut f\"{u}r Festk\"{o}rperforschung, Heisenbergstrasse 1, D-70569 Stuttgart, Germany}
\author{B. Keimer}
\affiliation{Max-Planck-Institut f\"{u}r Festk\"{o}rperforschung, Heisenbergstrasse 1, D-70569 Stuttgart, Germany}
\author{M. Le Tacon}
\affiliation{Karlsruhe Institute of Technology, Institut f\"{u}r Festk\"{o}rperphysik, D-76021 Karlsruhe, Deutschland}

\date{\today}

\begin{abstract}
We report on the effects of hydrostatic pressure (HP) on the charge density wave observed in underdoped cuprates. We studied YBa$_2$Cu$_3$O$_{6.6}$ (\textit{T$_c$}=61 K) using high-resolution inelastic x-ray scattering (IXS), and reveal an extreme sensitivity of the phonon anomalies related to the charge density wave (CDW) order to HP. The amplitudes of the normal state broadening and superconductivity induced phonon softening at Q$_{CDW}$ rapidly decrease as HP is applied, resulting in the complete suppression of signatures of the CDW below $\sim$1 GPa. Additional IXS measurements on YBa$_2$Cu$_3$O$_{6.75}$ demonstrate that this very rapid effect cannot be explained by pressure-induced modification of the doping level and highlight the different role of external pressure and doping in tuning the phase diagram of the cuprates. Our results provide new insights into the mechanisms underlying the CDW formation and its interplay with superconductivity.
\end{abstract}



\maketitle
The complex phase diagrams of correlated electron systems are shaped by the interplay between almost degenerate electronic states. In the high-$T_c$ superconducting cuprates, charge density wave (CDW) modulations have been ubiquitously observed, appearing as a generic feature of the moderately doped CuO$_2$ plane~\cite{Tranquada1995,Ghiringhelli2012,Chang2012,Comin2014,daSilvaNeto2014,Tabis2014,Croft2014,daSilvaNeto2015,Peng2017}.
While the doping and magnetic field dependence of the CDW order and its competition with superconductivity have been thoroughly studied~\cite{Ghiringhelli2012,Chang2012,Blanco2012,Blanco2014,Huecker2014}, the actual impact of CDW on the superconducting properties remains unclear. Recent studies of stripe ordered La$_{1.875}$Ba$_{0.125}$CuO$_4$ showed that the application of pressure increases \textit{T$_c$}, while drastically suppressing static spin and charge orders, as well as the local tilts of the CuO$_6$ octahedra which act as stripe pinning centers~\cite{Guguchia2013,Fabbris2013}. However, no direct measurements of the effect of pressure on the CDW in the other families of cuprates have been reported to date.

High pressures have been extensively used as a tuning parameter of the superconducting properties of the cuprates, and yielded the highest $T_c$ ever reported in these materials~\cite{Gao_PRB94,Monteverde2005} (and that held the world record until very recently, when it got surpassed by highly pressurized H$_3$S~\cite{Drozdov2015}).
In YBa$_2$Cu$_3$O$_{6+x}$, the few available structural studies revealed that hydrostatic pressure (HP) brings the superconducting CuO$_2$ planes and the Cu-O chains closer together, leading to 
a net increase of the number of holes per Cu atom in the CuO$_2$ planes~\cite{Jorgensen1990}. 
Recently it was shown that HP up to $\sim$ 30 GPa applied on fully oxygenated YBa$_2$Cu$_3$O$_{7}$, provides access to the so-far unreachable highly overdoped and non-superconducting phase of this material~\cite{Alireza2017}.
On the other hand, in the underdoped region of the phase diagram and particularly in the region of the $T_c$ vs doping plateau where the CDW is best developed, a large increase of $T_c$ is observed under HP, yielding $T_c$ largely exceeding that of the optimally doped compound.
The effective HP dependence of \textit{T$_c$} reflects the combination of pressure-induced doping with other HP effects, on \textit{e.g.} superexchange interactions, 
or oxygen ordering~\cite{Sadewasser2000,Struzhkin2016}. 
Given the competition between CDW and superconductivity in YBa$_2$Cu$_3$O$_{6+x}$, one might generally expect the CDW to be suppressed as $T_c$ increases under HP. The confirmation of this hypothesis and the measurement of the stability range of the CDW under pressure are important challenges in research on high-$T_c$ superconductivity. Answering these questions would also shine new light on whether the CDW can be held responsible for the $T_c$ vs doping $p$ plateau  of YBa$_2$Cu$_3$O$_{6+x}$~\cite{Liang2006}, or whether the competition between the CDW and superconductivity can account alone for the $dT_c/dP$ reported on this plateau~\cite{Cyr2015}. 

To address these issues, direct measurements of the pressure dependence of the CDW are needed, yet remain technically demanding. Resonant soft x-ray scattering which has been widely used to study the CDW is for instance not compatible with high pressure environments. Experiments with hard x-rays are impeded by serious difficulties arising from the weakness of the diffuse CDW features in underdoped YBa$_2$Cu$_3$O$_{6+x}$ ($\sim$10$^{-6}$ weaker than the (2 0 0) Bragg reflection~\cite{Chang2012}) in combination with the Compton scattering from the diamonds of a pressure generating diamond anvil cell.
 
To overcome these challenges, we turned towards high resolution inelastic x-ray scattering (IXS). Earlier investigations on underdoped YBa$_2$Cu$_3$O$_{6.6}$ have revealed two striking phonon anomalies in the IXS spectra around the CDW wave vector $Q_{CDW}$, which can be used to unambiguously determine the presence of the CDW. These are: i) a pronounced  broadening of the acoustic phonon for $T_c < T < T_{CDW}$ and ii) a strong softening of the low energy phonons below $T_c$~\cite{Blackburn2013b, LeTacon2013}. 

\begin{figure*}
\includegraphics[width=0.90\textwidth]{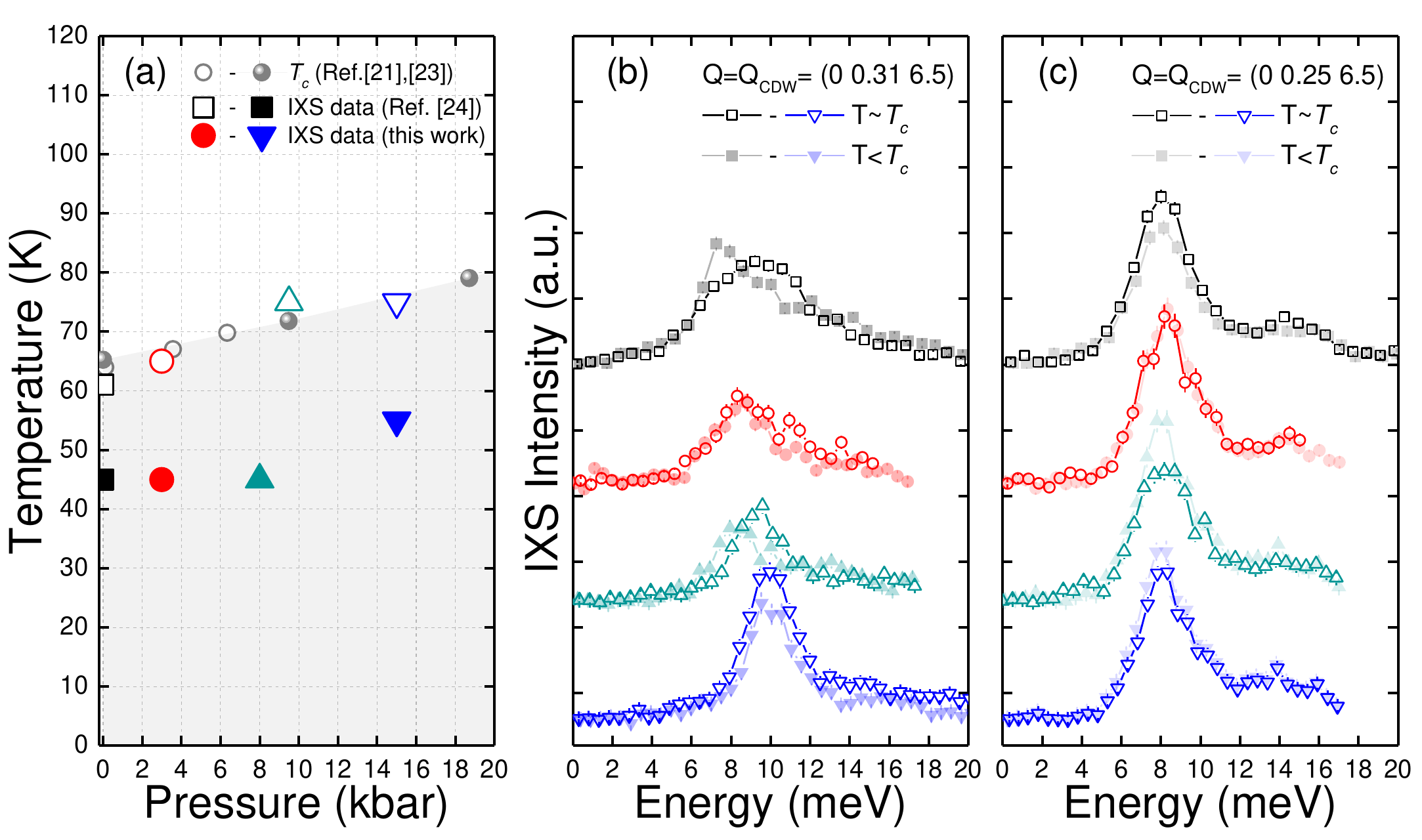}
\caption{(a) (\textit{P}, \textit{T}) phase diagram of YBa$_2$Cu$_3$O$_{6.6}$. The colored (black) symbols indicate the (\textit{P}, \textit{T}) points investigated by IXS in this work (Ref.~\cite{LeTacon2013}). The gray symbols show the superconducting transition temperature, \textit{T$_c$}, taken from resistivity and susceptibility data of Refs.~\cite{Sadewasser2000,Cyr2015}. (b,c) Pressure dependence of the inelastic part of the IXS spectra spectra of YBa$_2$Cu$_3$O$_{6.6}$ at Q=Q$_{CDW}$=(0 0.31 6.5) (b) and Q=(0 0.25 6.5) (c), recorded at $\sim$\textit{T$_c$} (open symbols) and below \textit{T$_c$} (solid symbols) and following the symbol/color code of panel (a). The intensities shown in these two panels were corrected for the Bose factor and the spectra are vertically shifted for clarity. The ambient pressure data are taken from Ref.~\cite{LeTacon2013}.} \label{Fig1}
\end{figure*}

Here, we monitor these anomalies above and below $T_c$ as function of applied HP in underdoped YBa$_2$Cu$_3$O$_{6.6}$. Our results reveal that the CDW-related phonon anomalies respond strongly to the application of external pressure, with their amplitude rapidly decreasing under pressure and resulting in their complete suppression below $\sim$1 GPa.
We further demonstrate that these effects cannot be solely attributed to pressure-induced doping effects, and conclude that the unusual pressure dependence of $T_c$ in this regime of doping cannot be fully attributed to presence of the CDW.


The experiments were performed on high quality, detwinned single crystals of ortho-VIII-ordered YBa$_2$Cu$_3$O$_{6.6}$ (\textit{p}=0.12, \textit{T$_c$}=61 K) and ortho-III ordered YBa$_2$Cu$_3$O$_{6.75}$ (\textit{p}=0.134, \textit{T$_c$}=71 K), grown by a flux method as described elsewhere~\cite{Lin2002}. The oxygen content was selected through an annealing procedure under the appropriate temperature and oxygen partial pressure conditions, and the hole doping level was determined based on the known doping dependence of the c-axis lattice parameter and of \textit{T$_c$}~\cite{Lindemer1989,Jorgensen1990b,Liang2006}. 
HP and low temperature conditions were applied using a gasketed diamond anvil cell (DAC) positioned in a custom-designed continuous-flow helium cryostat. The pressure was varied in-situ using a helium pressurized membrane. Optimal hydrostatic conditions were obtained using compressed helium as the pressure transmitting medium. The pressure was calibrated using the ruby luminescence method~\cite{Syassen2008}. 
The IXS measurements were performed at the ID28 beamline of the ESRF with an incident photon energy of 17.794 keV and a corresponding instrumental energy resolution of 3 meV. Throughout the paper, the momentum transfers are quoted in reciprocal lattice units (r.l.u.) of the orthorhombic crystal structure. The momentum resolution within the scattering plane was set to $\sim$0.016 (0.048) r.l.u. along the b$^*$ (c$^*$) direction. The x-ray beam was focused to a 50 $\mu$m $\times$ 40 $\mu$m spot on the sample surface. 

\begin{figure}
\includegraphics[width=0.90\linewidth]{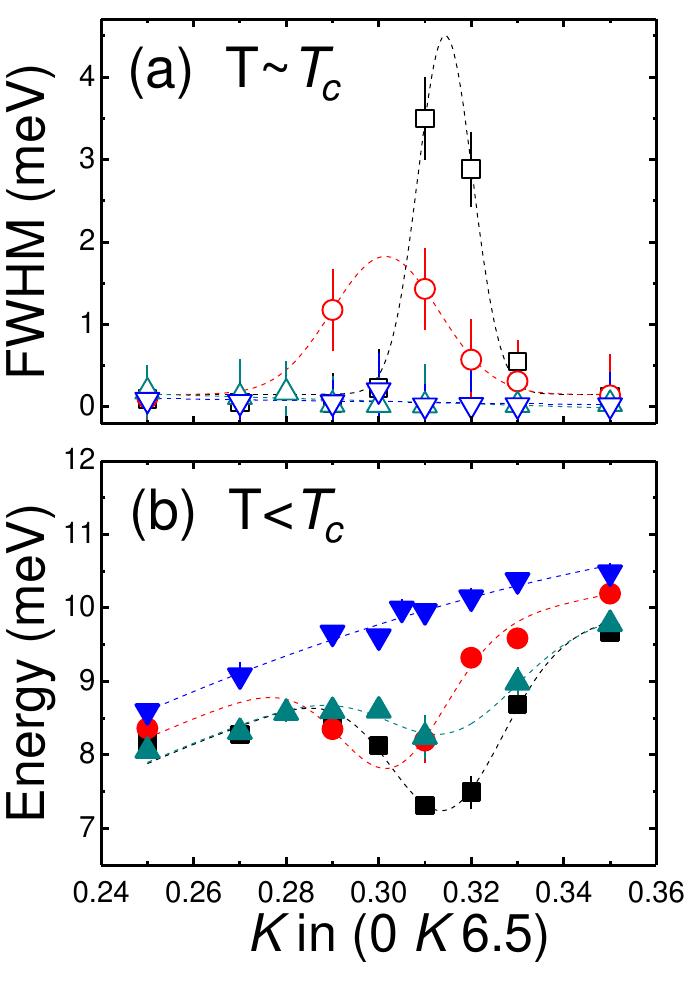}
\caption{Pressure and momentum dependence of the acoustic phonon linewidth (FWHM) at $\sim$\textit{T$_c$} (a) and the phonon energy below \textit{T$_c$} (b). The dashed lines are guides to the eyes. The data follow the symbol/color code of the phase diagram of Fig.~\ref{Fig1}(a).} \label{Fig2}
\end{figure}

Following our earlier ambient pressure measurements, high pressure IXS spectra of YBa$_2$Cu$_3$O$_{6.6}$ were recorded in the Brillouin zone adjacent to the (0 0 6) Bragg reflection, across the [010] direction and close to the strong (0 0.31 6.5) CDW superstructure peak, probing transverse acoustic and optical phonons of the B$_1$ representation~\cite{LeTacon2013}.
The pressure dependence of the low energy phonon spectra at Q$_{CDW}$=(0 0.31 6.5) and at Q=(0 0.25 6.5) at $\sim$\textit{T$_c$} and below \textit{T$_c$} is plotted in Fig.~\ref{Fig1}.
The recorded spectra were fitted using standard damped harmonic oscillator functions convoluted with the experimental resolution. The results of these fits are given in Fig.~\ref{Fig2}.

We first focus on the pressure dependence of the phonon behavior close to \textit{T$_c$} (the pressure dependence of \textit{T$_c$} is given in Fig.~\ref{Fig1}(a)). The CDW-related acoustic phonon broadening observed at ambient pressure in a narrow momentum range around Q$_{CDW}$, is clearly observed in the data recorded at 3 kbar, albeit strongly reduced in amplitude: while at 1 bar the transverse acoustic phonon linewidth (FWHM) reaches 3.5 meV at Q$_{CDW}$, at 3 kbar it is reduced to 1.5 meV (Fig.~\ref{Fig2}(a)). On further increasing pressure to 9.5 kbar the phonon linewidth at Q$_{CDW}$ becomes resolution-limited, as it is also away from Q$_{CDW}$ for all the pressure and temperature values measured (Fig.~\ref{Fig1}(c)). 
Within our experimental resolution, and in agreement with the ambient pressure data, we do not observe any anomalies in the phonon energies close to \textit{T$_c$}.
We note here that the momentum dependence of the elastic intensity in the IXS spectra, even though it becomes harder to resolve in DAC measurements, appears featureless at 9.5 kbar and is therefore in line with the pressure dependence of the phonon broadening. 

We now turn to the IXS spectra recorded in the superconducting state, where under ambient pressure conditions the acoustic phonon linewidths are resolution limited and the phonon dispersions exhibit a pronounced dip around Q$_{CDW}$, with a softening of $\sim$1.3 meV upon cooling from \textit{T$_c$}=61 K to 45 K. While the softening is already reduced by 50 $\%$ at 3 kbar, the dispersion dip at Q$_{CDW}$ is clearly observed up to 7 kbar (Fig.~\ref{Fig2}(b)). Under a HP of 15 kbar no anomaly is observed in the dispersion curve. The phonon linewidths in the superconducting state remained resolution limited at Q$_{CDW}$ and away from it.

The IXS data demonstrate that all CDW-related phonon anomalies observed at ambient pressure disappear at a modest pressure of less than $\sim$1 GPa. 
Earlier HP studies on YBa$_2$Cu$_3$O$_{6.6}$ have determined a pressure-induced effective doping of the CuO$_2$ planes in the range of $\sim$0.008-0.017 holes/GPa~\cite{Jorgensen1990,Almasan1992,Zhang1995,Gupta1995,Cyr2015}.

Doping dependent x-ray diffraction data on YBa$_2$Cu$_3$O$_{6+x}$ have however shown that the CDW intensity and correlation length are comparable for \textit{p}=0.12 and \textit{p}=0.134~\cite{Blanco2014,Huecker2014}, indicating that the suppression of the phonon anomalies under HP cannot be attributed to a doping-induced suppression of the CDW.
Nevertheless, given that the detailed doping dependence of the CDW-related phonon anomalies in YBa$_2$Cu$_3$O$_{6+x}$ has not been reported, the potential disappearance of the CDW anomalies for \textit{p}>0.12 cannot be excluded based on these data alone.

\begin{figure}
\includegraphics[width=0.99\linewidth]{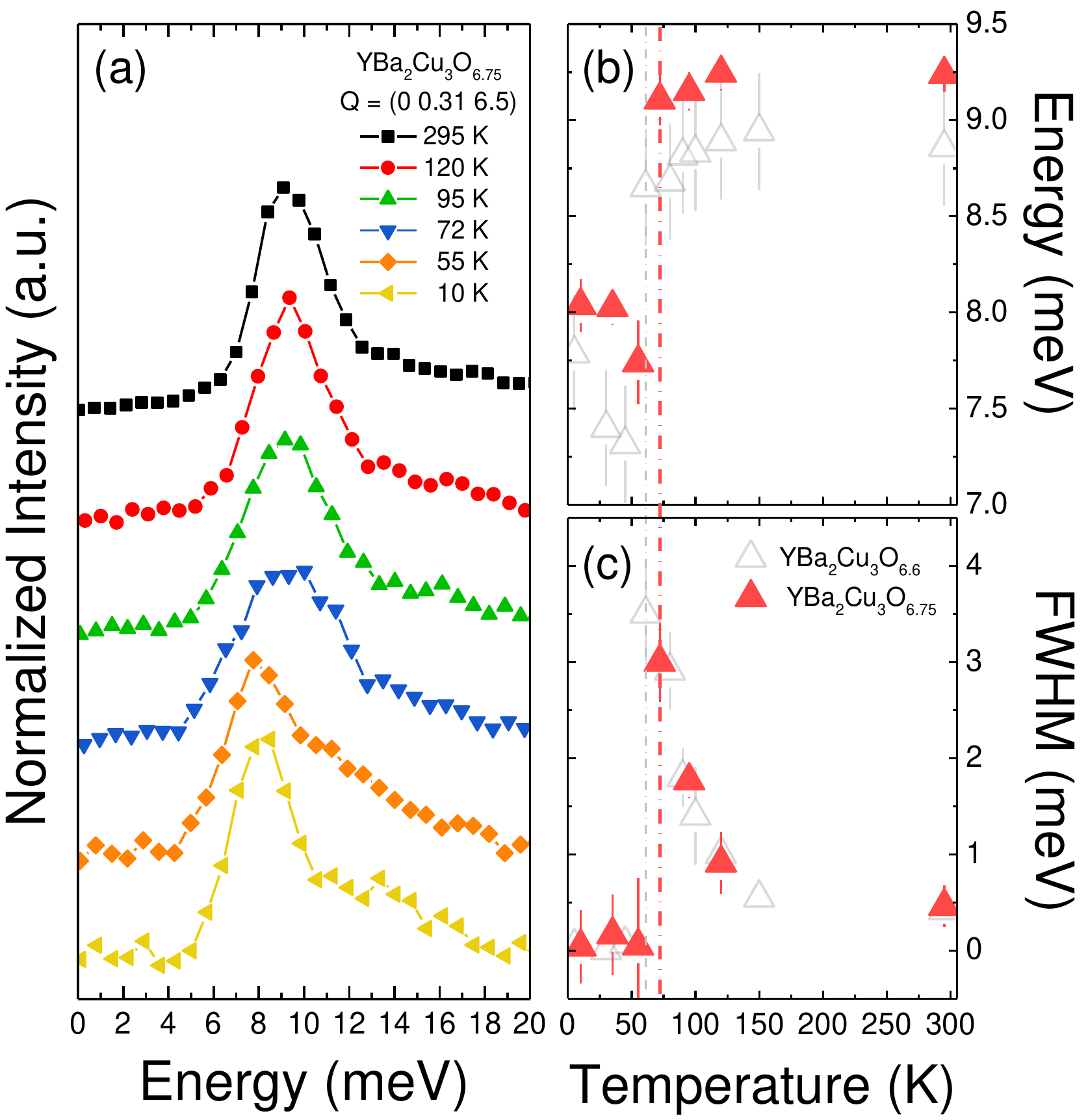}
\caption{(a) Temperature dependence of the inelastic part of the IXS spectra of YBa$_2$Cu$_3$O$_{6.75}$ at Q = Q$_{CDW}$= (0 0.31 6.5). The spectra were corrected for the Bose factor, and were vertically shifted for clarity. (b, c) Temperature dependence of the acoustic phonon energy (b) and FWHM (c), at Q = Q$_{CDW}$= (0 0.31 6.5) for YBa$_2$Cu$_3$O$_{6.6}$ (grey symbols) and YBa$_2$Cu$_3$O$_{6.75}$ (red symbols).} \label{Fig3}
\end{figure}

To test this scenario we performed IXS measurements on YBa$_2$Cu$_3$O$_{6.75}$ (\textit{p}=0.134), with $T_c$ = 71 K, very close to that of YBa$_2$Cu$_3$O$_{6.6}$ under $\sim$1 GPa (see also Fig.~\ref{Fig1}(a)). The results are summarized in Fig.~\ref{Fig3}, and clearly show that the low temperature anomalies observed in both the linewidth and the energy of the acoustic phonon for \textit{p}=0.12~\cite{LeTacon2013} are also present for \textit{p}=0.134. 
This demonstrates that the suppression of the phonon anomalies in YBa$_2$Cu$_3$O$_{6.6}$ cannot simply be attributed to pressure-induced doping.
The present data on YBa$_2$Cu$_3$O$_{6.75}$, together with those previously reported on YBa$_2$Cu$_3$O$_{6.55}$~\cite{Blackburn2013b}, YBa$_2$Cu$_3$O$_{6.6}$ and fully oxygenated YBa$_2$Cu$_3$O$_{7}$~\cite{LeTacon2013} clearly show that the doping dependence of the phonon anomalies tracks that of the CDW peak as obtained from diffraction data.
The absence of phonon anomalies above $\sim$1 GPa therefore implies a HP-induced suppression of the CDW order.


Having ruled out a  pressure-induced modification of the doping level as the origin of the rapid pressure dependence of the CDW, we now discuss alternative scenarios. The strong competition between CDW order and superconductivity, which was previously inferred from temperature and magnetic field dependent measurements~\cite{Chang2012,Blanco2012}, could weaken the CDW if the intrinsic strength of the pairing interaction and/or the interlayer Josephson coupling increased under pressure. 
Indeed, as in the case of light-induced transient superconductivity recently reported~\cite{Kaiser_PRB2014,Hu_NatureMaterials2014} and observed up to temperatures significantly larger than that of the CDW (which also rapidly disappears with the THz pumping~\cite{Foerst_PRB2014}), this indicates that the details of the structural modifications under pressure, and of the subsequent changes in the microscopic interaction schemes between electrons might play a key role~\cite{Mankowsky_Nature2014}.
For instance, Raman scattering experiments have shown that the antiferromagnetic exchange interaction \textit{J}, one of the candidates for the electronic pairing, increases under HP application~\cite{Aronson1991,Eremets1991,Struzhkin2016}.  It is also worth mentioning that according to earlier resistivity measurements the pseudogap and its onset temperature follow the increase of \textit{T$_c$} under HP, a result which was linked to a common intrinsic pressure dependence of superconductivity and the pseudogap~\cite{deMello2002,Solovjov2016}.


Regarding possible structural origins of the enhancement of superconductivity and the destabilization of the CDW under pressure, we note that early HP neutron diffraction experiments on YBa$_2$Cu$_3$O$_{6.6}$ have revealed an enhanced compression of the apical oxygen O(4) - planar copper Cu(2) bond length~\cite{Jorgensen1990}, which is considered to play a key role in the charge transfer mechanism and is accurately reproduced by recent bond valence sum calculations~\cite{Gao2017}.  
In terms of linear compressibilities under hydrostatic conditions, at $\sim$1 GPa the a (b) axis was found to change by -0.24$\%$ (-0.21$\%$), whereas the more compressible c axis changes by -0.44$\%$~\cite{Jorgensen1990}.
Given however that the available structural data are limited to pressures smaller than 6 kbar, experimental studies of the atomic parameters in YBa$_2$Cu$_3$O$_{6+x}$ for an extended pressure range are required  to allow a quantitative understanding of the HP effects. This information would furthermore provide input for calculations of the Fermi surface topology under HP, that has been suspected to determine some of the CDW features~\citep{Comin2014}.   
However, the existing electronic structure calculations report significant changes in the Fermi surface only for compressions much higher than the ones corresponding to the suppression of the CDW-related anomalies reported here~\cite{Khosroabadi2002,Sidorov2015}. Changes in the Fermi-arc electronic structures at pressures as modest as those used here to suppress the CDW are therefore expected to be marginal. This indicates that the existence of the Fermi arcs alone may not be sufficient to drive the system towards a CDW instability. Related conclusions have been drawn from the recent observation of a robust CDW state in overdoped (Bi,Pb)$_{2.12}$Sr$_{1.88}$CuO$_{6+\delta}$, a compound without Fermi surface nesting features~\cite{Peng2017}.

Next, we discuss the implications for the Fermi surface reconstruction revealed by quantum oscillations (QOs). Small electron pockets arising from a CDW-induced Fermi surface reconstruction have been suggested as the origin of the low frequency QOs observed in underdoped cuprates~\cite{Sebastian2015}. There have not been such measurements under HP to date, except for recent ones in underdoped stoichiometric YBa$_2$Cu$_4$O$_8$~\cite{Putzkee2016}. In these experiments, QOs are observed up to pressures of $\sim$8.5 kbar, with an increased frequency, consistent with increased doping and decreased $Q_{CDW}$. One should however be careful when comparing this with results obtained in YBa$_2$Cu$_3$O$_{6+x}$ as, to date, no direct evidence for the CDW has been reported in YBa$_2$Cu$_4$O$_8$, although several indirect indications (including the QOs) are pointing towards their existence. Given the consistent evolution of $T_c$ with pressure in the two systems, one would likely expect similar disappearance of the CDW in pressurized YBa$_2$Cu$_4$O$_8$. At this stage, this can only be reconciled with the observation of the QOs arguing that the large magnetic field required for these experiments 'resurrects' the CDW once superconductivity has been sufficiently weakened. Searching for the CDW in YBa$_2$Cu$_4$O$_8$ and investigating QOs under high pressure in YBa$_2$Cu$_3$O$_{6+x}$ are now required to settle this issue.

We end our discussion remarking that the increase of $T_c$ under HP application on YBa$_2$Cu$_3$O$_{6.6}$ is the net outcome of different uniaxial pressure derivatives. 
Experiments under uniaxial strain have indeed shown that c-axis compression, which brings the CuO$_2$ planes closer together, induces a charge transfer and an enhancement of $T_c$, while b- (a-) axis strain, which increases (reduces) the structural orthorhombicity of YBa$_2$Cu$_3$O$_{6.6}$, results in an increase (decrease) of $T_c$~\cite{Meingast1991,Welp1992,Hardy2010}. 
Uniaxial pressure experiments, which modify the crystal and electronic structure selectively and continuously~\cite{Pickett1997,Pickett1997b}, are therefore needed to complement our hydrostatic experiments. 

In summary, we have presented high quality IXS data on YBa$_2$Cu$_3$O$_{6.6}$ under HP, following the low temperature phonon linewidth and energy anomalies at Q$_{CDW}$. 
The CDW-related phonon anomalies show an extreme fragility to the application of external pressure and disappear already below $\sim$1 GPa, suggesting the pressure-induced suppression of the CDW order. 
Parallel measurements on YBa$_2$Cu$_3$O$_{6.75}$ exclude pressure-induced charge transfer as the origin of the CDW suppression and highlight the different effect of chemical doping and pressure-induced structural modifications in the underdoped cuprates. Instead, pressure-induced strengthening of superconductivity and/or subtle changes in the Fermi surface geometry may be responsible for this effect. In any case, the simultaneous application of pressure and magnetic fields opens up new opportunities for the quest to understand the origin of CDW formation and its influence on the transport properties of the cuprates.



\end{document}